\documentclass[twocolumn,prb]{revtex4}
\usepackage{graphicx,amsmath}

\begin{document}

\title {Superconducting proximity effect through graphene from zero field to the Quantum Hall regime.}
\author{Katsuyoshi Komatsu, Chuan Li, S. Autier-Laurent, H. Bouchiat and  S. Gu\'eron}
\affiliation{Laboratoire de Physique des Solides, Univ. Paris-Sud, CNRS, UMR 8502, F-91405 Orsay Cedex, France.}

\today

\begin{abstract}
We investigate the superconducting proximity effect through graphene in the long diffusive junction limit, at low and high magnetic field. The interface quality and sample phase coherence lead to a zero resistance state at low temperature, zero magnetic field, and high doping. We find a striking suppression of the critical current near graphene\rq{}s charge neutrality point, which we attribute to specular reflexion of Andreev pairs at the interface of charge puddles. This type of reflexion, specific to the Dirac band structure, had up to now remained elusive. At high magnetic field the use of superconducting electrodes with high critical field enables the investigation of the proximity effect in the Quantum Hall regime. Although the supercurrent is not directly detectable in our two wire configuration, interference effects are visible which may be attributed to the injection of Cooper pairs into edge states.

\end{abstract}

\maketitle

\section{Introduction}
The celebrated electronic  band structure of graphene leads to many interesting features. Among them is the possibility to tune its carrier density from electron to hole, with the consequence that the Integer Quantum Hall effect is observed over a wide range of magnetic fields. Another consequence is the fact that transport can proceed via carriers of either the conduction or the valence band, depending on the doping, and may even proceed via a conversion of one type of carrier into the other, across regions of different doping \cite{Klein} the so called Klein tunneling effect.

\begin{figure} 
\begin{center} 
\includegraphics[clip=true,width=7cm]{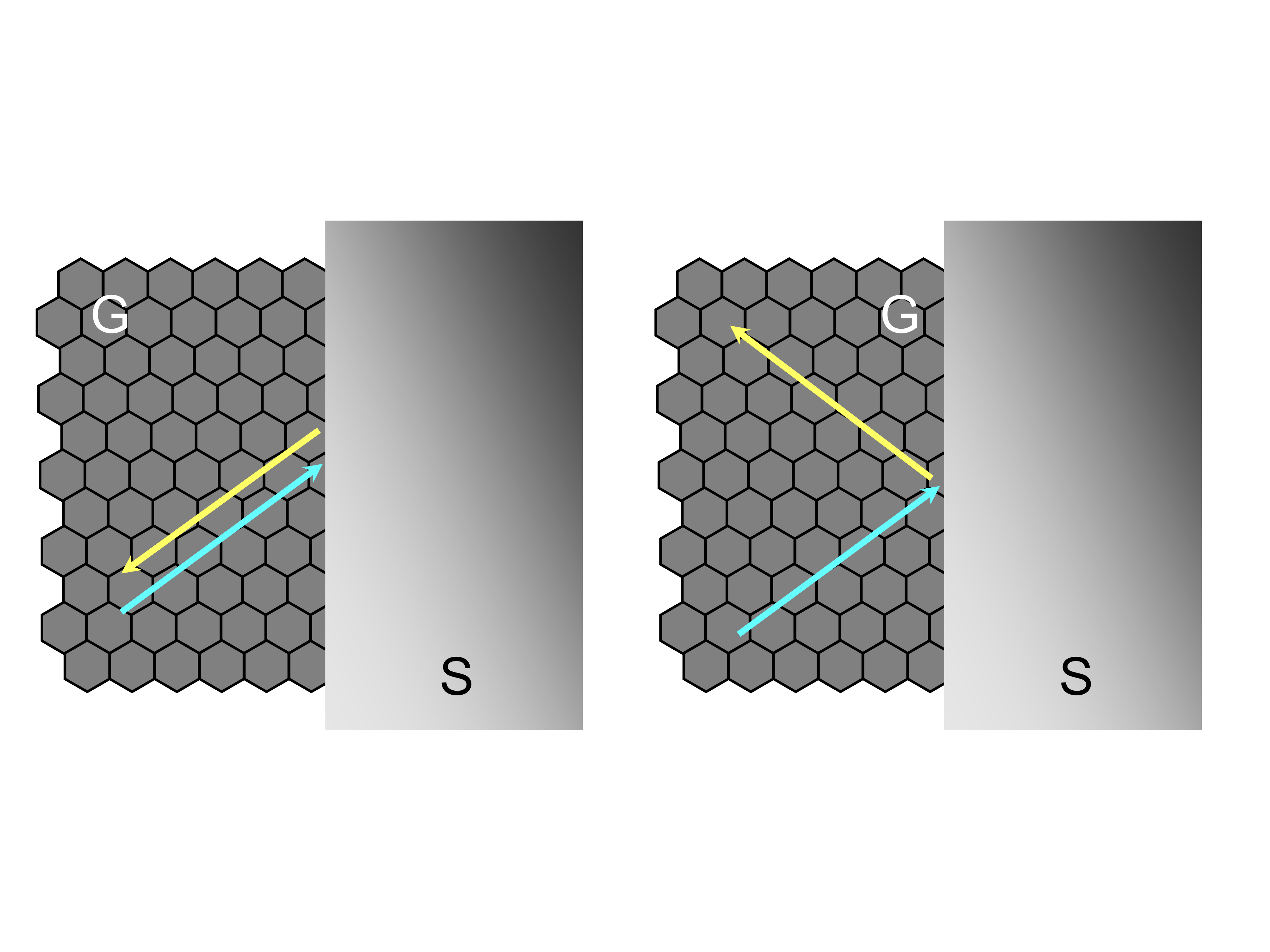}
\end{center}
\caption{Sketch of the retro- and specular Andreev reflexion at a G/S interface. Left: Retro-reflexion occurs in usual conductors and in doped graphene, where the Fermi energy much exceeds the superconducting electrode\rq{}s energy gap $\Delta$, $E_F\gg\Delta$. Right: The specular Andreev reflexion occurs in graphene at doping small enough that $E_F\ll\Delta$.}
\label{AR} 
\end{figure}
\begin{figure} 
\begin{center} 

\centerline{\includegraphics[clip=true,width=7cm]{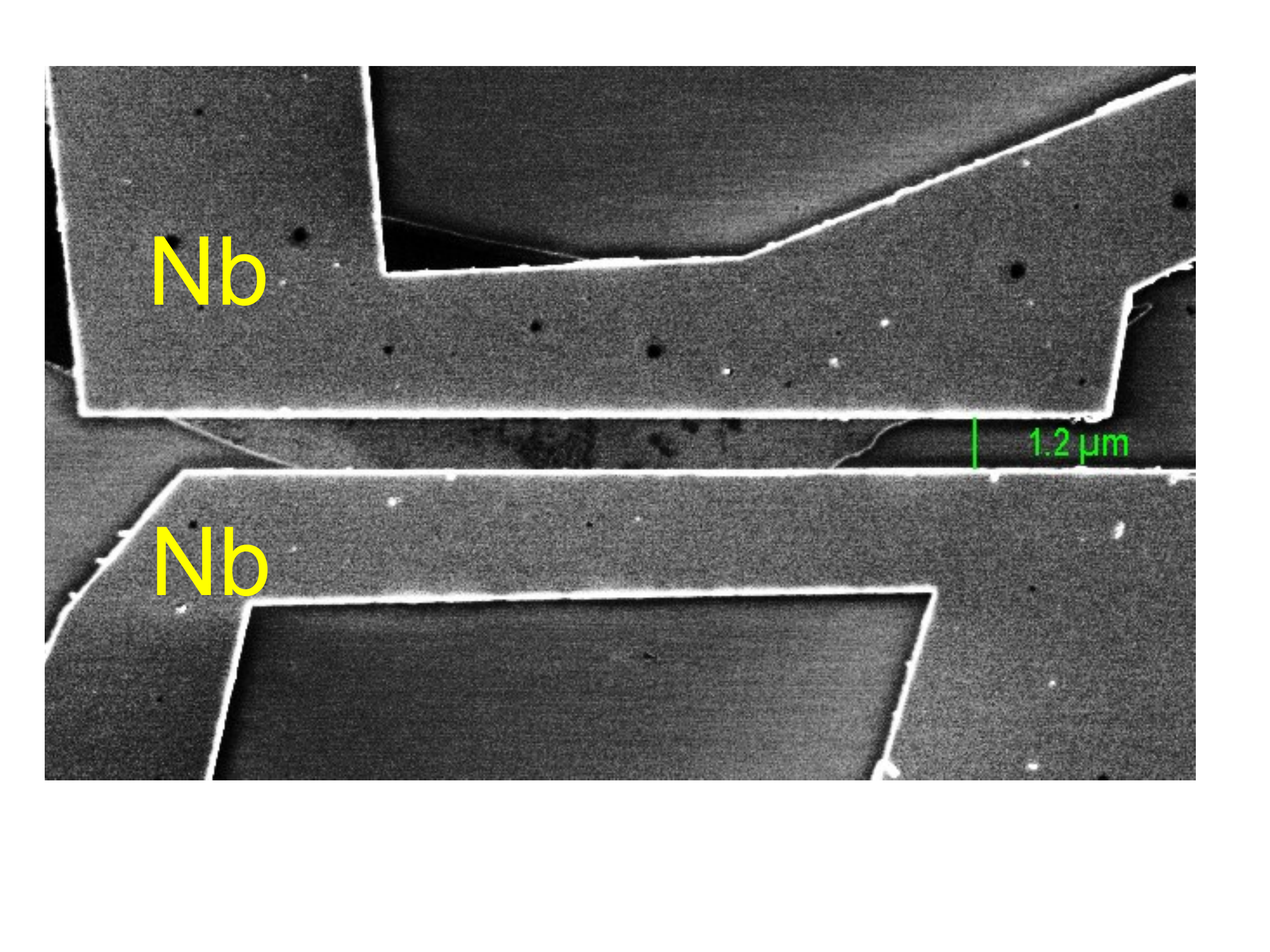}}
\end{center}
\caption{Scanning electron micrograph of the graphene sample connected to Nb electrodes. The distance between electrodes is $L=1.2$~$\mu$m and the graphene width is $W=12$~$\mu$m.}
\label{photo} 
\end{figure}
It was suggested \cite {Beenakker} that a superconductor/graphene interface should also reveal the fact that the valence and conduction band touch at the so called Dirac point. Indeed, transport across a Superconductor/Normal metal (S/N) interface at subgap energy implies extracting two electrons from the superconductor and injecting them into the N, which produces a correlated Andreev pair in the normal metal. In a usual normal metal, which is highly doped in the sense that the Fermi level lies well within the conduction band, both electrons are injected in the conduction band of the N. The two injected members of the Andreev pair then follow the same, albeit time-reversed, diffusive path in the normal conductor, so that coherent propagation can occur over several micrometers (the phase coherence length at low temperature). This coherent propagation leads to supercurrents that flow through such normal conductors several microns long connected to two superconductors. In contrast, at a superconductor/graphene (S/G) interface, if the superconductor\rq{}s Fermi level is aligned with the graphene Dirac point, the two electrons of a Cooper pair must split into an electron in the conduction band and the other in the valence band. The two members of the injected pair in the graphene now have the same velocity (rather than opposite) parallel to the S/G interface (see Fig. \ref{AR}) and thus do not follow the same diffusive path. The observation of this special type of pair injection, also called \lq\lq{}specular Andreev reflexion\rq\rq{}, has so far remained elusive. This is because the doping inhomogeneities in the graphene samples, of several millielectronvolts \cite{Martin}, are much larger than the  superconductor\rq{}s energy gap. Thus only the usual injection of counter-propagating electron pairs (also called Andreev retroreflexion) sets in. 

In this article we show that diffusive transport of Andreev pairs through quantum coherent graphene reveals an analog of specular Andreev reflexion at an S/G interface, in the form of specular reflexions of Andreev pairs at the interface between a doped charge puddle and a zero density region. These processes result in the destruction of counter-propagation upon specular reflexion, and lead to a large phase accumulation withing each Andreev pair. Since all pairs contribute to the supercurrent with their phase, the resulting supercurrent is suppressed. We argue that this specular reflexion explains the suppression of the critical current that we observe near the Charge Neutrality Point (CNP) in our quantum coherent, long and diffusive SGS junctions.

In the second part of the article, we explore the possibility of injecting Cooper pairs in graphene in the Quantum Hall regime. In contrast to the low field proximity effect, the supercurrent is no longer carried by many diffusing pairs, but must be carried exclusively by the chiral edge states. Thus the two injected electrons must propagate on opposite edges of the graphene sheet. We present a long SGS junction which sustains a tunable supercurrent at low magnetic field. The superconducting electrodes, made of a high critical field superconductor, remain superconducting at fields such that the graphene exhibits integer quantum Hall plateaus, indicating that transport proceeds via edge states. We present non linear transport features which hint to the existence of interference, controlled by gate voltage or magnetic field, between the electrons propagating along different edges of the graphene.

\section{Long junction samples in low magnetic field}

\subsection{Sample fabrication}
Several SGS junctions were fabricated, in which the length of graphene between S electrodes was greater than one micron, more than twice as long as previously reported \cite{Heersche,Andrei,Ojeda,Finkelstein,Doh}. Such lengths place a great constraint on the sample in order for a full proximity effect to develop: the phase coherence length must be longer than the sample length, and the interface quality must be excellent since a low transparency decreases the critical current through the junction\cite{Cuevas}. The critical current itself, in the case of a perfect interface, scales as the inverse length cubed (see discussion further down). 
In addition, the temperature must be low since the critical current is roughly exponentially suppressed by temperature with a coefficient proportional to the diffusion time across the sample, which scales as the square of the sample length \cite{Dubos}.

Thus it is not surprising that not all samples we fabricated showed a full superconducting proximity effect at low temperature. Out of 12 samples with superconducting electrodes (of different superconducting materials and contact layers), 3 exhibited a full proximity effect when cooled to low temperature.
All samples were mechanically exfoliated with the tape method and deposited on a doped silicon substrate previously cleaned in an oxygen plasma. Standard electron-beam lithography was performed and the contacts were sputtered onto the samples after an hour long annealing step in vacuum at $100^{\circ}{\rm C}$. The contacts consist of a thin Pd layer, 4 to 8 nm thick, over which the superconducting layer, either Nb or ReW \cite{Raffy}, is deposited without breaking vacuum, with a thin Pd cover layer. 
We report in this article results on a SGS junction consisting of a $W=12~\mu$m-wide graphene sheet with a $L=1.2~\mu$m separation between Nb electrodes (Fig.~\ref{photo}). A second junction, with ReW electrodes, is $W=2.6~\mu$m-wide with a length of $L=0.7~\mu$m between electrodes.
The samples are tested at room temperature and then thermally anchored to the mixing chamber of a dilution refrigerator, and measured via low pass filtered lines.
\subsection{Critical current in zero magnetic field}

\begin{figure} 
\begin{center} 
\includegraphics[clip=true,width=8cm]{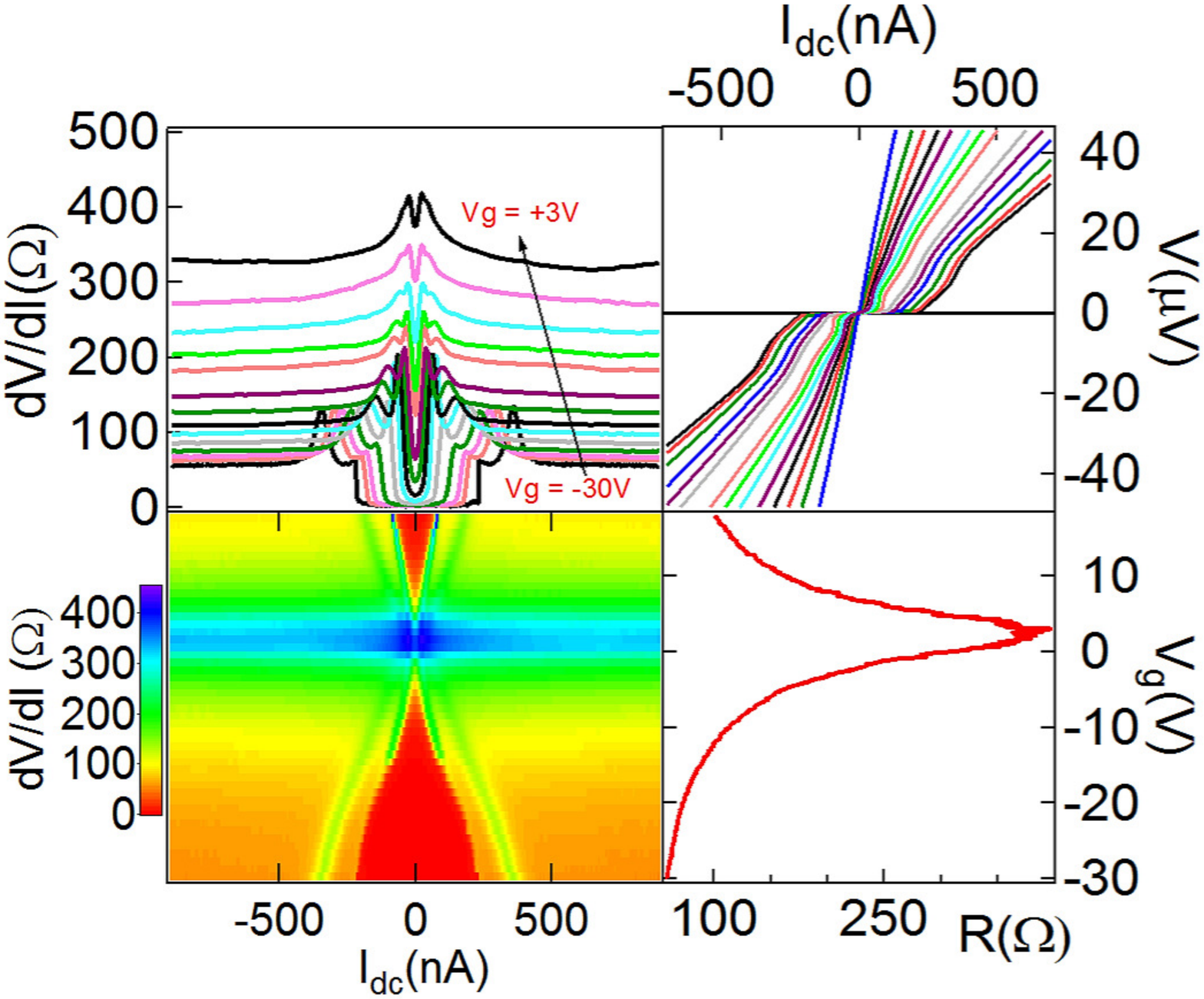}
\end{center}
\caption{Proximity effect in  graphene connected to Nb electrodes at 200 mK. Upper left panel: $dV/dI$ vs $I_{\rm dc}$ for different galte voltages, and, bottom left panel, its 2 dimensional color plot. The suppression of critical current in a gate voltage region of $\pm 10V $ around the charge neutrality point is noticeable. Upper right panel: I(V) curves for different gate voltages, showing how the proximity effect varies between a full proximity effect with zero resistance at high doping, and quasi normal behavior with a linear IV around the charge neutrality point.  Lower right panel:  Zero bias differential resistance as a function of gate voltage in the normal state, from which the $R_N$ is determined. A small magnetic field was applied to destroy the constructive interference leading to the supercurrent.}
\label{Nb_Scurrent} 
\end{figure}

\begin{figure} 
\begin{center} 
\includegraphics[clip=true,width=9cm]{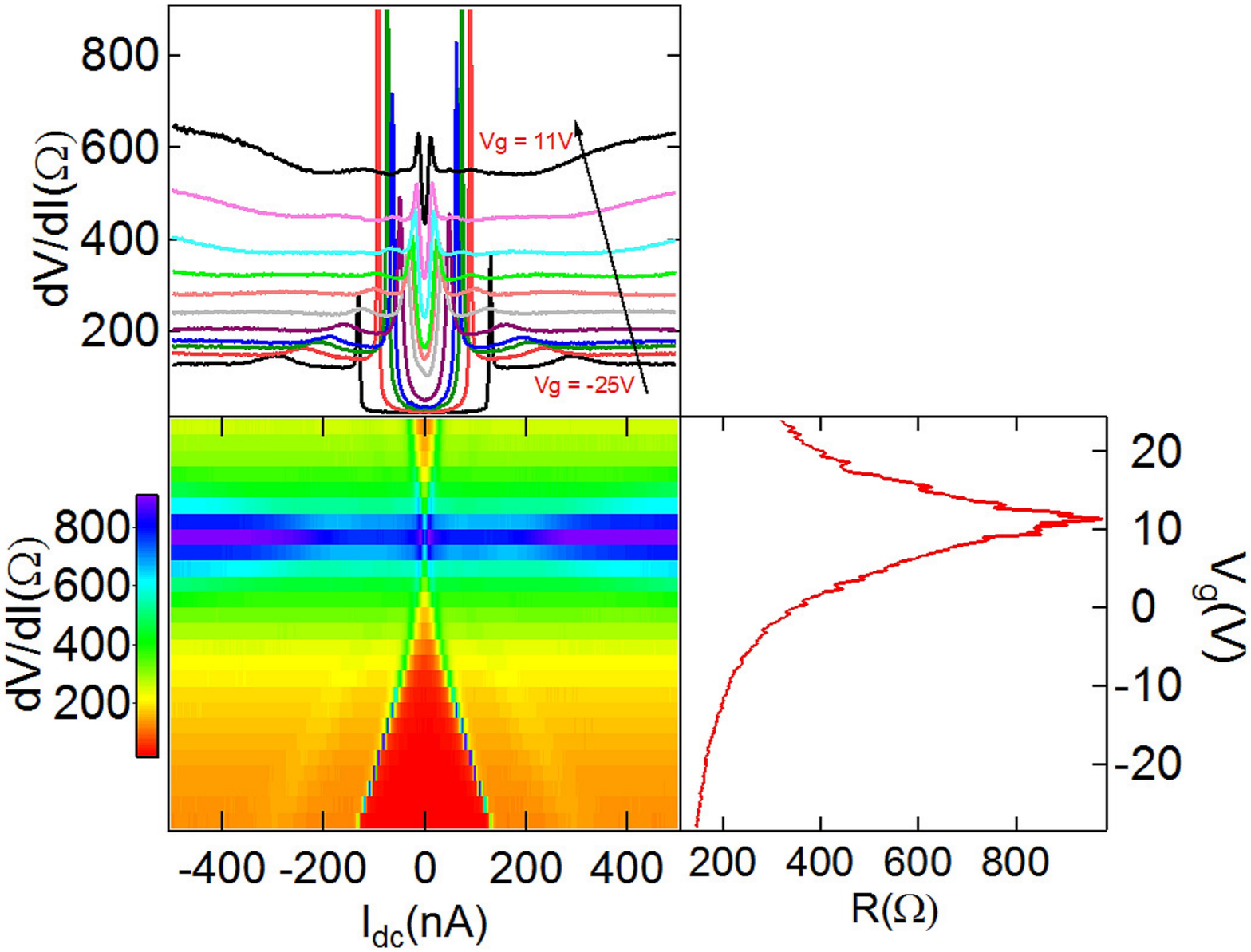}
\end{center}
\caption{Proximity effect in graphene connected to ReW electrodes at 55 mK. Top panel: $dV/dI$ vs $I_{\rm dc}$. Bottom left panel: 2 dimensional colour plot emphasizing the suppression of the supercurrent around the charge neutrality point. Right panel: Resistance as a function of gate voltage in a small magnetic field which suppresses the constructive interference leading to supercurrent.}
\label{ReW_Scurrent} 
\end{figure}

\begin{figure} 
\begin{center} 
\includegraphics[clip=true,width=8.5cm]{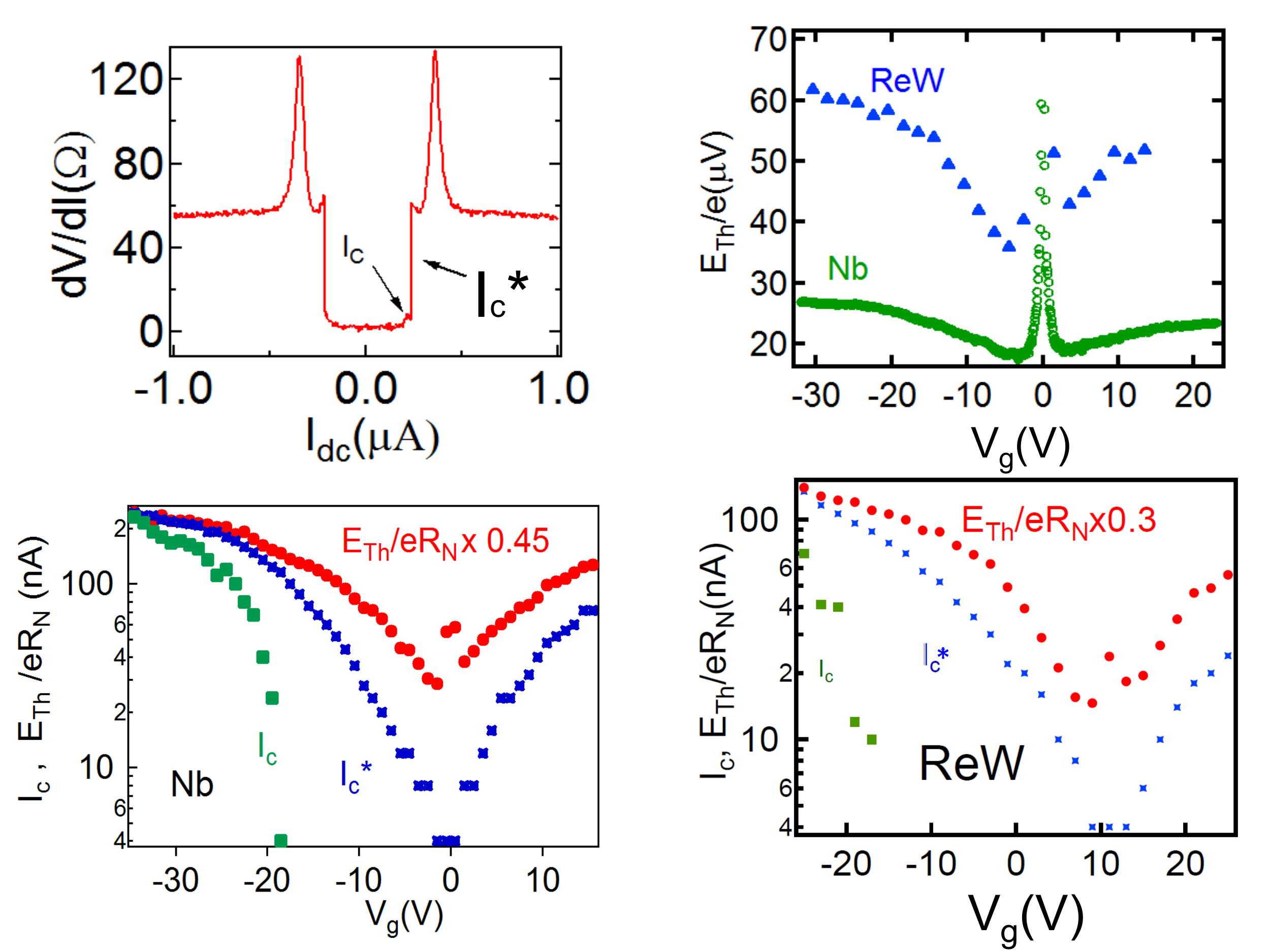}
\end{center}
\caption{Comparison of switching current with Thouless energy. Upper left panel: Two ways of defining the switching current: $I_{\rm c}$, the largest current for which the differential resistance $dV/dI$ is zero, and $I_{\rm c}^*$, the inflection point of the jump in $dV/dI$ towards large resistance. Upper right panel: Variations of the Thouless energy with gate voltage, deduced from the sample resistance in the normal state, for both samples. The resistance of the Nb sample was measured at 1K. The resistance of the ReW sample was measured at 55 mK at a current bias above the critical current of the proximity effect. Bottom panels: Comparison of $\displaystyle I_{\rm c}$ and $\displaystyle I_{\rm c}^*$ with $E_{\rm Th}/eR_N$ for the sample with Nb electrodes at 200 mK, and with ReW electrodes at 55 mK.}
\label{Ic_ETh} 
\end{figure}

Both SGS junctions display a gate tunable supercurrent at low temperature, as shown in Fig.~\ref{Nb_Scurrent} and Fig. \ref{ReW_Scurrent}. As is clear in the figures, the critical current is strongly suppressed near the charge neutrality point, and we argue that this suppression is due to the specular reflexion at the charge puddle interfaces. To quantify this suppression, we compare the measured critical current (assumed here to be equal to the switching current, the current at which the junction resistance switches from zero to a finite value) to the theoretically expected critical current (see Fig. \ref{Ic_ETh} for the definition of the critical current). In the theory of the proximity effect in the diffusive, long junction limit, the critical current  has a maximum zero temperature value given by the Thouless energy $E_{ Th}$  divided by the normal resistance state $R_N$, multiplied by a numerical factor $\alpha$ which depends on the junction length $L$:  $I_{c} =\alpha E_{Th}/eR_{N}$, where $E_{Th}=\hbar D/L^2$, with  $D=v_Fl_e/2$ the diffusion constant in two dimensions, $v_F$ the Fermi velocity and $l_e$ the mean free path. 

The tunability of graphene is an asset to probe this relation. As shown in Fig.~\ref{Ic_ETh}, one can compare the measured switching current to the Thouless energy divided by the normal state resistance as the gate voltage is varied. It is clear from the figure that there is not a constant factor between $E_{Th}/eR_{N}$ and  $I_{c}$ but that  $I_{c}$ is strongly suppressed  at small gate voltage, as the charge neutrality point is approached.

This suppression has not been reported in the other graphene based SNS junctions, which are more than two or three times shorter than the devices reported in this article.

\begin{figure} 
\begin{center} 
\includegraphics[clip=true,width=8cm]{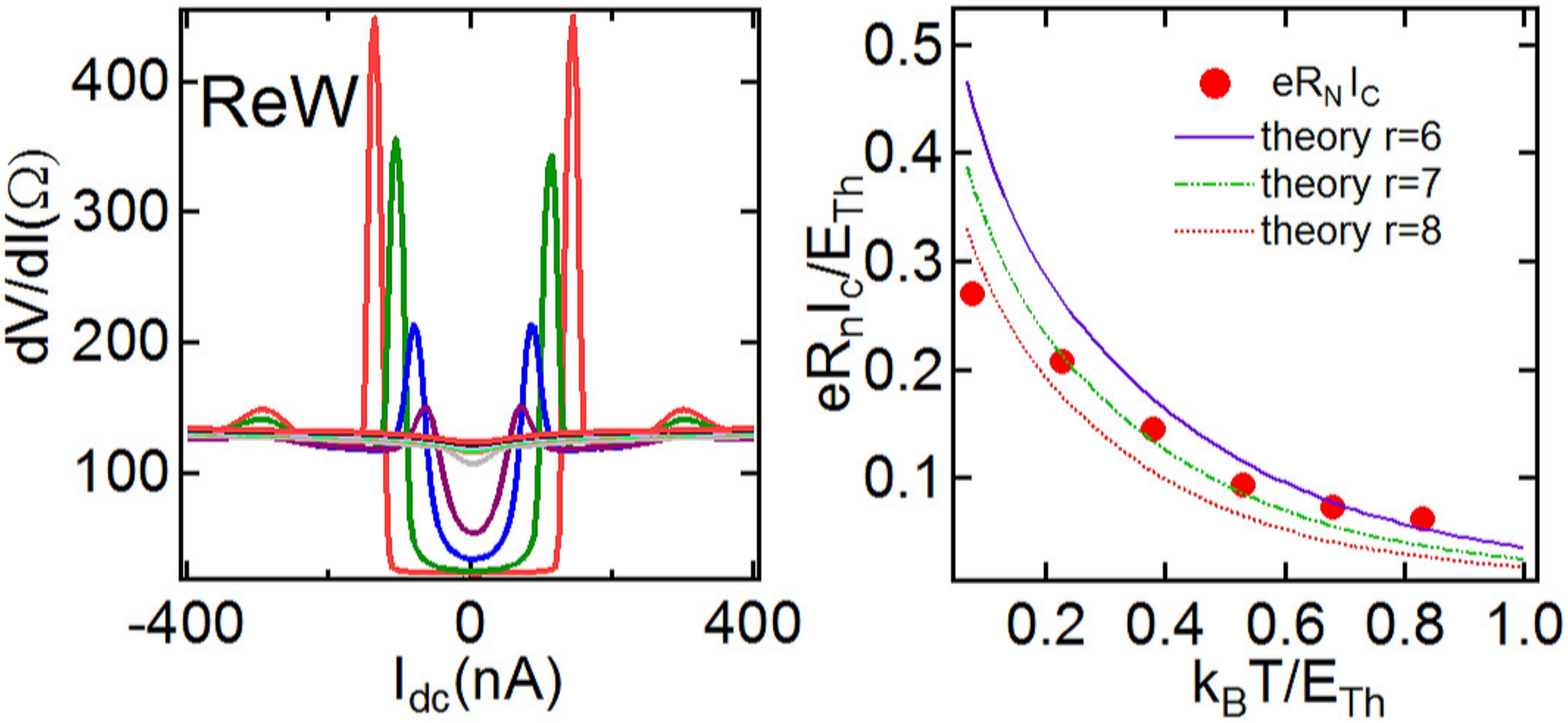}
\end{center}
\caption{Temperature dependence of the proximity effect through the ReW sample. Left panel, differential resistance curves at temperatures ranging from 100 mK to 800 mK. Right panel, comparison of the extracted critical currents with the theoretically expected decay with temperature, for different ratios $r$ of the contact resistance to the graphene sheet resistance. Both the overall suppression of the critical current with respect to the Thouless energy at low temperature, and the decay with temperature, are accounted for assuming a ratio $r$ of roughly 7.  }
\label{Ic_T} 
\end{figure}

To interpret the data of Fig. \ref{Ic_ETh}, we first discuss the maximum critical current and its temperature dependence (Fig. \ref{Ic_T}), which we explain by a non ideal interface. We then address the gate-voltage induced suppression of the critical current.
The maximum critical current to be expected depends on how long the junction is, compared to the superconducting coherence length in the graphene layer, defined as $\xi_s=\sqrt{\hbar D/\Delta}$, with $\Delta$ the electrode\rq{}s superconducting gap, and $D$ the diffusion constant in graphene. We find that $L/\xi_s=5$ for the ReW sample and 7 for the Nb one, which places these junctions in the long (but not infinitely long) junction limit. As computed in \cite{Dubos}, this gives an expected coefficient $\alpha$ between $I_{c}$ and $E_{Th}/eR_{N}$ of 9 and 8 at zero temperature for the Nb and ReW samples respectively, close to the 10.8 value of the infinitely long junction. These theoretical values are more than twenty times larger than the maximum measured $\alpha$ coefficient of 0.5 for Nb and 0.3 for ReW. This reduced critical current is a feature noted in practically all experiments on S/graphene/S junction, and is attributed to partial transmission at the S/graphene interface. The temperature dependence of the critical current (Fig. \ref{Ic_T}) confirms the partial transmission of the interface, since the critical current decay with temperature is faster than expected for a perfect interface, as described in \cite{Cuevas}. Fig. \ref{Ic_T} shows the variations of the differential resistance curves with temperature, as well as the comparison of the critical current suppression with theoretical prediction considering an opaque interface. From the comparison one can extract a rather large interface resistance, roughly five times larger than the resistance of the graphene sheet itself.

We now argue that the critical current suppression near the CNP cannot be attributed to finite temperature. The effect of temperature is twofold: First, the thermal fluctuations induced by $k_BT$ must be smaller than the Josephson coupling $E_J=\frac{\hbar}{2e}I_c$, which gives a minimal supercurrent of  44 nA/K. Thus the minimal critical current at the experimental temperatures are 9 nA at 200mK and 2 nA at 50 mK, and do not depend on gate voltage.
Second, temperature decreases the switching current in a manner that is predicted by the Usadel equations, that has been numerically solved exactly \cite{Dubos}, and that can be approximated by an exponential decay as $I_c(T)\approx I_c(0)e^{-T/10E_{Th}}$ for a perfect interface and $I_c(T)\approx I_c(0)e^{-T/3E_{Th}}$ for an opaque interface. Since the overall variation with gate voltage of the Thouless energies of both samples (deduced from the measured resistance R via the diffusion constant $\displaystyle D=\frac{L}{W}\frac{1}{ne^2R}$ with $n$ the carrier density) is less than a factor $50\%$ (between 20 and 30 $\mu eV$ for the Nb sample, and between  40 and 60 $\mu eV$ for the ReW sample, see Fig. \ref{Ic_ETh}), it cannot explain the gate-voltage induced suppression by a factor 10. Thus it is clear that the remarkable suppression of the supercurrent near the Charge Neutrality Point cannot be explained by temperature-induced effects.

\begin{figure} 
\begin{center} 
\includegraphics[clip=true,width=7cm]{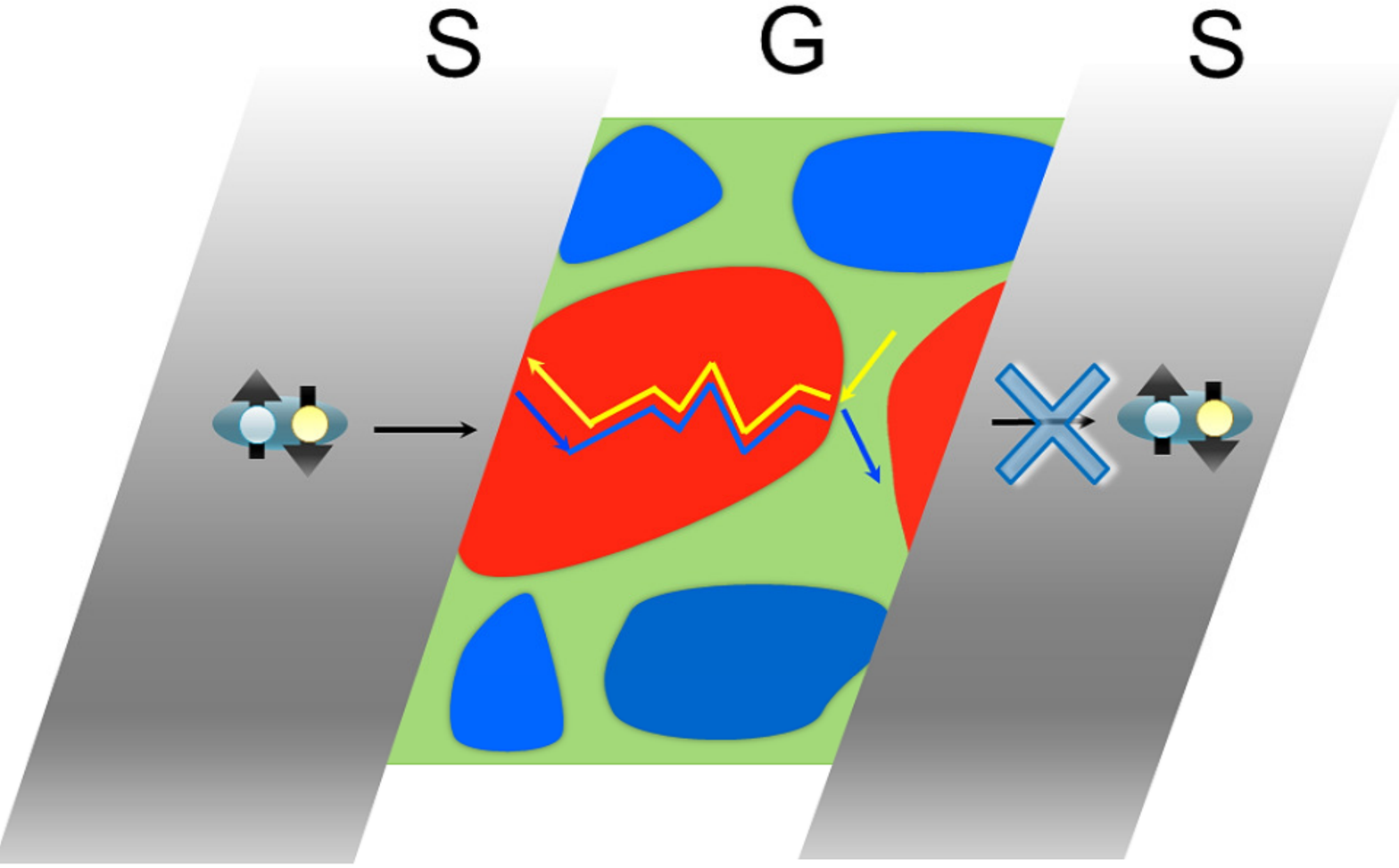}
\end{center}
\caption{Sketch of how the specular reflexion of an Andreev pair at an n/0 junction can lead to loss of counterpropagation and thus large phase accumulation within an Andreev pair. The red region is electron doped, the blue one hole doped, and the green region in between has zero doping.}
\label{speculardiffusion} 
\end{figure}

We attribute this suppression close to the CNP to specular reflection of an Andreev pair at the charge puddle contours, as sketched in Fig. \ref{speculardiffusion}.  Indeed, around the CNP, electron-doped regions coexist with hole doped ones, forming a network of so called puddles \cite{Martin}. Where the doping varies from n to p doping there is necessarily a boundary with exactly zero doping, to within $k_BT$, termed a 0 region. Thus a time-reversed Andreev pair formed by the usual Andreev retroreflexion at the superconductor/graphene interface has, near the CNP, a large probability of encountering a n/0 or p/0 boundary. At such boundaries such junctions, a specular-like reflexion must occur when two counter propagating electrons  diffusing in the n-doped region are converted into two electrons belonging to two different bands in the 0 region. The change in relative velocity destroys the counter propagation of the pair. As the two electrons diffuse across the rest of the graphene, they undergo uncorrelated scattering events and their relative phase difference increases. Since the total supercurrent is the sum of all contributions from the propagating Andreev pairs, constructive interference is destroyed when counter-propagation is lost, and thus the supercurrent is suppressed (see fig. \ref{speculardiffusion}). Interestingly, the effect of these puddles is immense in the superconducting state (and presumably all the more so that the superconducting coherence length, the \lq\lq{}size\rq\rq{} of the pair, is small with respect to the puddle size), whereas it is much weaker in the normal state where thanks to Klein tunneling, the puddles do not suppress single quasiparticle propagation so much.

In summary, whereas the specular Andreev reflexion in ballistic S/G/S junctions can yield a supercurrent \cite{Titov}, we have shown that in diffusive S/G/S junctions a specular-like reflexion of Andreev pairs at p/0 or n/0 junctions leads to accumulation of phase difference within the Andreev pair. The critical current is then suppressed, in a manner which depends on the number of such n/0 (or p/0) junctions within the sample. This translates into a critical current suppressed most near the charge neutrality point. The supercurrent suppression by charge puddles is thus expected to be largest in samples that are long (large ratio of sample length $L$ to puddle size, typically larger than 50 nm \cite{Martin}) and connected to superconductors with large gaps, corresponding to smaller superconducting coherence lengths ($\xi_s=\sqrt{\hbar D/\Delta}$ is typically 125 nm in graphene for Nb ($\Delta=1.6$~meV) or 170~nm for ReW ($\Delta=1.2~meV$, as compared to 350~nm for Al ($\Delta=0.2$~meV), given the diffusion constant $D=4.10^{-2}m^2/s$ in these graphene samples.

\subsection{Junction under radiofrequency irradiation}
\begin{figure} 
\begin{center} 
\includegraphics[clip=true,width=8cm]{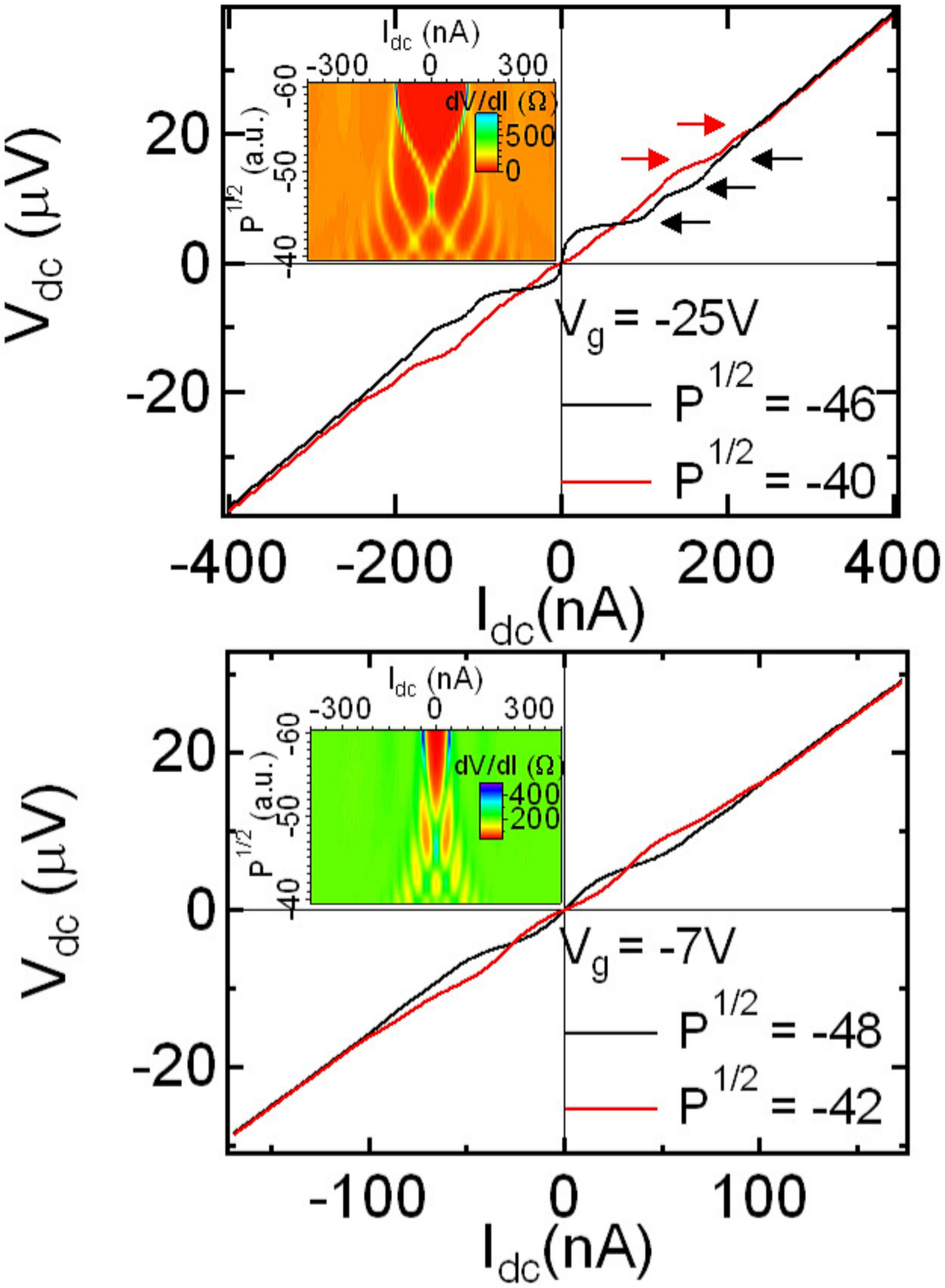}
\end{center}
\caption{Effect of radiofrequency irradiation on the junction with ReW electrodes. Top panel: Junction under irradiation of 2.4 GHz, at a gate voltage of $V_g=-25V$. Bottom panel: junction under irradiation of 2.4 GHz, at a gate voltage of $V_g=-7V$ for which no full proximity effect (supercurrent) is observed, but only a lower low bias differential resistance. Nonetheless Shapiro like features develop under irradiation. The insets display the rf power dependence of the proximity effect.  The arrows point to the dc voltage plateaus, distant by $\Delta V=5.3~\mu V$, close to the expected interval $\Delta V=\hbar \omega /2e=4.9~ \mu V$.}
\label{Shapiro} 
\end{figure}

As also reported by others, the junctions display Shapiro steps, i.e. replica of the zero resistance state, which appear at finite dc voltage, when submitted to radiofrequency irradiation (via an antenna placed near the sample). This is shown in the top panel of Fig. \ref{Shapiro} for the sample with ReW electrodes, at high doping, which displays a full proximity effect with a critical current of 130~nA. What is more original is the observation of sequential non linearities in the IV curves of the junctions at gate voltages such that a full proximity effect with a zero resistance state does not develop, demonstrating that non linearities in the IV curve are sufficient to induce phase locking and replica of non linear features (bottom panel).

\subsection{Suppression of supercurrent by small magnetic field}
\begin{figure} 
\begin{center} 
\includegraphics[clip=true,width=8cm]{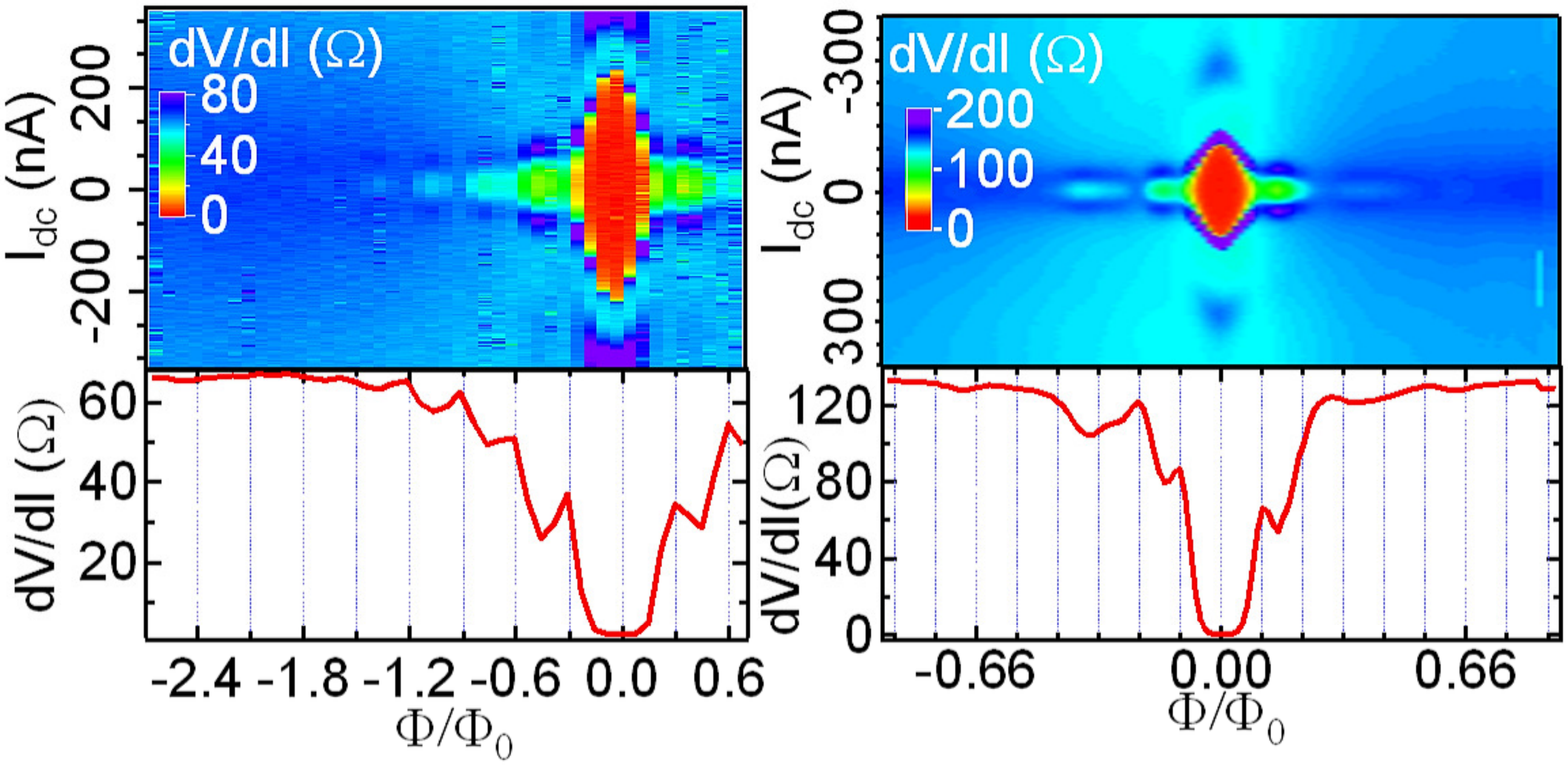}
\end{center}
\caption{Low field dependence of $dV/dI(I)$ for the sample with Nb (left) and ReW (right) electrodes, at T=200 mK for Nb and 55 mK for ReW, and at high doping. Bottom panels: Line traces of $dV/dI$ at zero current bias as a function of magnetic flux through the graphene. We attribute the small period of the flux dependence to strong focusing of the magnetic field by the large superconducting electrodes.}
\label{Fh_Nb} 
\end{figure}

Fig. \ref{Fh_Nb} displays the differential resistance as a function of current for different magnetic fields, and shows that the supercurrent is suppressed  in an oscillatory manner, as expected for wide proximity junctions \cite{FhCuevas,Francesca}. However the supercurrent is not recovered periodically, but rather the resistance oscillates away from zero in a periodic manner. We attribute the absence of full supercurrent recovery to the asymmetric (trapezoid-like) shape of the graphene samples, and to probable irregularities in the transmission between electrodes and graphene, which lead to inhomogeneous supercurrent densities \cite{Barone}. The fact that the oscillation period is smaller than one flux quantum $\Phi_0$ through the sample is attributed to the focusing effect of the field by the superconducting electrodes. Although the interference patterns look similar for both samples, one can notice an asymmetry in the field dependence of the sample with ReW electrodes, which we attribute to trapped flux in these high $H_{c2}$ but low $H_{c1}$ electrodes.

\section{Proximity effect in the integer Quantum Hall regime}
The observation of supercurrent through graphene contacted to the high $H_c$ superconductor ReW (see previous part) suggests the exciting possibility of observing a supercurrent running through a conductor in the Quantum Hall regime. Such a supercurrent would have to be carried by edge states, so that the time reversed electrons injected from the superconductor would be injected into the edge states at the opposite edges of the sample. Only few authors have considered this scenario theoretically \cite{Ma,Stone}. They have shown that in principle such a proximity effect is possible in the integer quantum Hall regime, with a maximal critical current given by the ballistic limit of $ev_d/L$, where $L$ is the perimeter of the sample and $v_d$ the drift velocity. In the following we show that we achieve the quantum Hall regime in graphene with superconducting electrodes, and present elements which suggest the existence of coherent interference within the sample, modulated by magnetic field or gate voltage, hinting to a tunable proximity effect through graphene in the quantum Hall regime.

\subsection{Integer Quantum Hall regime}
\begin{figure} 
\begin{center} 
\includegraphics[clip=true,width=9cm]{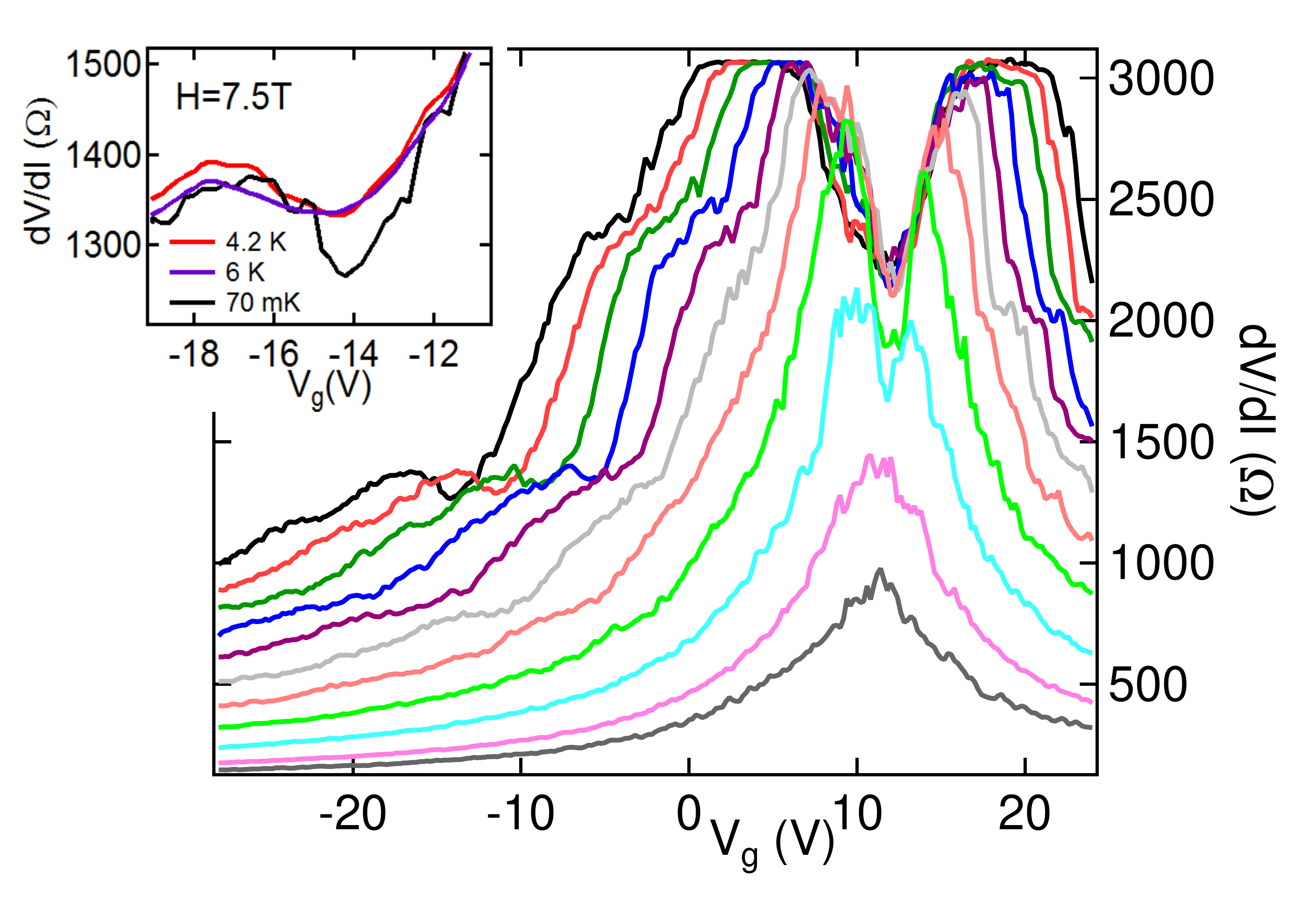}
\end{center}
\caption{Two wire differential resistance as a function of gate voltage for the sample with ReW ($H_c>7.5~T$), at magnetic fields from 0 to 7.5 T, every Tesla between 0 and 5~T, and every 0.5 Tesla above 5~T. Temperature is 70 mK. The inset displays how the Hall plateau at 7.5 T  and $V_g=-14~V$ flattens out as temperature is increased.}
\label{QHE_ReW} 
\end{figure}

\begin{figure} 
\begin{center} 
\includegraphics[clip=true,width=8cm]{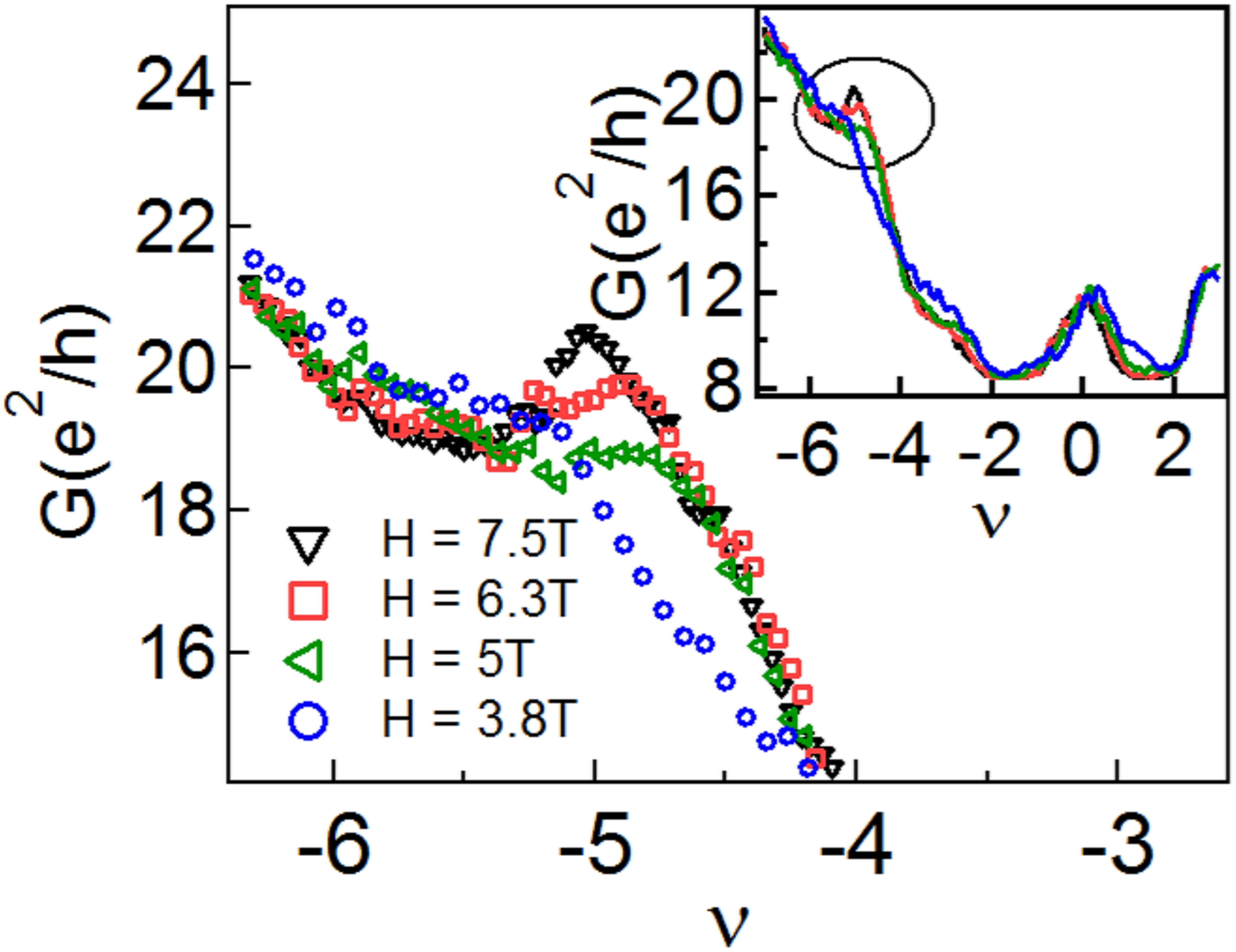}
\end{center}
\caption{Quantum Hall effect of graphene sample with ReW electrodes, plotted as a function of filling factor $\nu = en/Bh$. Inset: Two wire differential conductance as a function of filling factor in the Quantum Hall regime. Main panel: zoom of the $\nu=-6$ region, which displays oscillations in conductance of up to $10\%$ at the edge of the plateau. }
\label{QHE_ReW_normalized} 
\end{figure}

Fig. \ref{QHE_ReW} displays the zero current differential resistance of the SGS junction as a function of gate voltage, for fields between 0 and 7.5 T, at low temperature (70 mK). The quantum Hall effect is visible, in the form of plateaus, at fields above 5 T. Indeed, it has been shown that the quantum Hall regime is detectable in a two wire measurement, in the form of regions in which the conductance is quantized at the Hall conductance value\cite{Marcus}. The exact shape of the conductance versus filling factor curve (i.e., whether peaks or dips separate the plateau regions) depends on the sample aspect ratio since the two wire resistance is a weighed combination of the sample\rq{}s $\rho_{xx}$ and $\rho_{xy}$  \cite{Marcus}.  Fig.~ \ref{QHE_ReW_normalized} shows that the filling factors corresponding to the plateaus are those expected for graphene ($\nu=nh/(eB)=\pm 2,\pm 6,...$), but that the values of the conductance plateaus are larger than those expected for graphene. We attribute this discrepancy to scattering, which broadens the Landau levels, and to sample inhomogeneities typical of wide graphene sheets, which change the plateaus conductance values, as has been observed by others \cite{Marcus}. The factor of almost 3 in conductance enhancement could also be interpreted as due to three  effective samples in parallel. We note that this lack of correct quantification is found in wide samples and not in square samples (see e.g. the quantum Hall regime in a different sample with Nb electrodes and a square shape, shown in the Appendix in Fig. \ref{QHE_Nb_both}).

\subsection{Proximity effect in the Quantum Hall regime}

The curves of Fig. \ref{QHE_ReW} show no obvious trace of supercurrent at high field in the form of a zero resistance state, even though the electrodes are superconducting up to more than 7.5 T (we found a critical current of $3.5~\mu A$ at 7.5 T and low temperature, measured through slightly wider ReW leads, see appendix). This is in fact to be expected,  given the two wire configuration which mixes the $\rho_{xx}$ and $\rho_{xy}$ components. It is actually interesting to ask what should be the signature of a supercurrent in the quantum Hall regime, especially measured in a two wire configuration.

Some hints of the superconducting proximity effect can be found, however. We show below that we find signatures of the proximity effect both in the incoherent regime, where the S/graphene/S junction can be viewed as two uncorrelated S/graphene junctions in series, and in the coherent regime, where signatures of the coherent propagation of pairs through the graphene via quantum Hall edge states are visible.

The incoherent proximity effect is visible in the shape of the plateaus themselves. As shown in the inset of Fig. \ref{QHE_ReW} for the $\nu=-6$ plateau, and also reported in AlGaAs/GaAs heterostructures connected to high $H_{c2}$ NbN electrodes \cite{Takayanagi}, the plateaus are far less flat at low temperature than at high temperature. The resistance at the transition between two Hall plateaus exhibits a non monotonous variation with filling factor, with a decrease of resistance of up to ten percent. This amplitude variation of the resistance was interpreted in \cite{Takayanagi} as the effect of a change in conductance at an NS interface with respect to an NN interface as the edge channel transmission coefficient changes with filling factor.
Analytical and numerical computations of the NS conductance in the specific case of the quantum Hall regime were considered in \cite{Takagaki,Schoen}. They predict that the NS conductance is not twice the NN conductance, in contrast to what one might naively expect for two electrons being transmitted via perfectly conducting edge channels at the Quantum Hall plateau. This is because the two electrons of a pair must travel along different edges, much as in the normal case. However interference effects at the NS interfaces lead to a predicted oscillatory behavior around the Quantized Hall conductance in \cite{Takagaki}. When disorder at the interface is included, \cite{Schoen} find that the two-wire conductance is at most the Quantized Hall value , in contrast to our experimental results and those of \cite{Takayanagi}.

\begin{figure} 
\begin{center} 
\includegraphics[clip=true,width=8.5 cm]{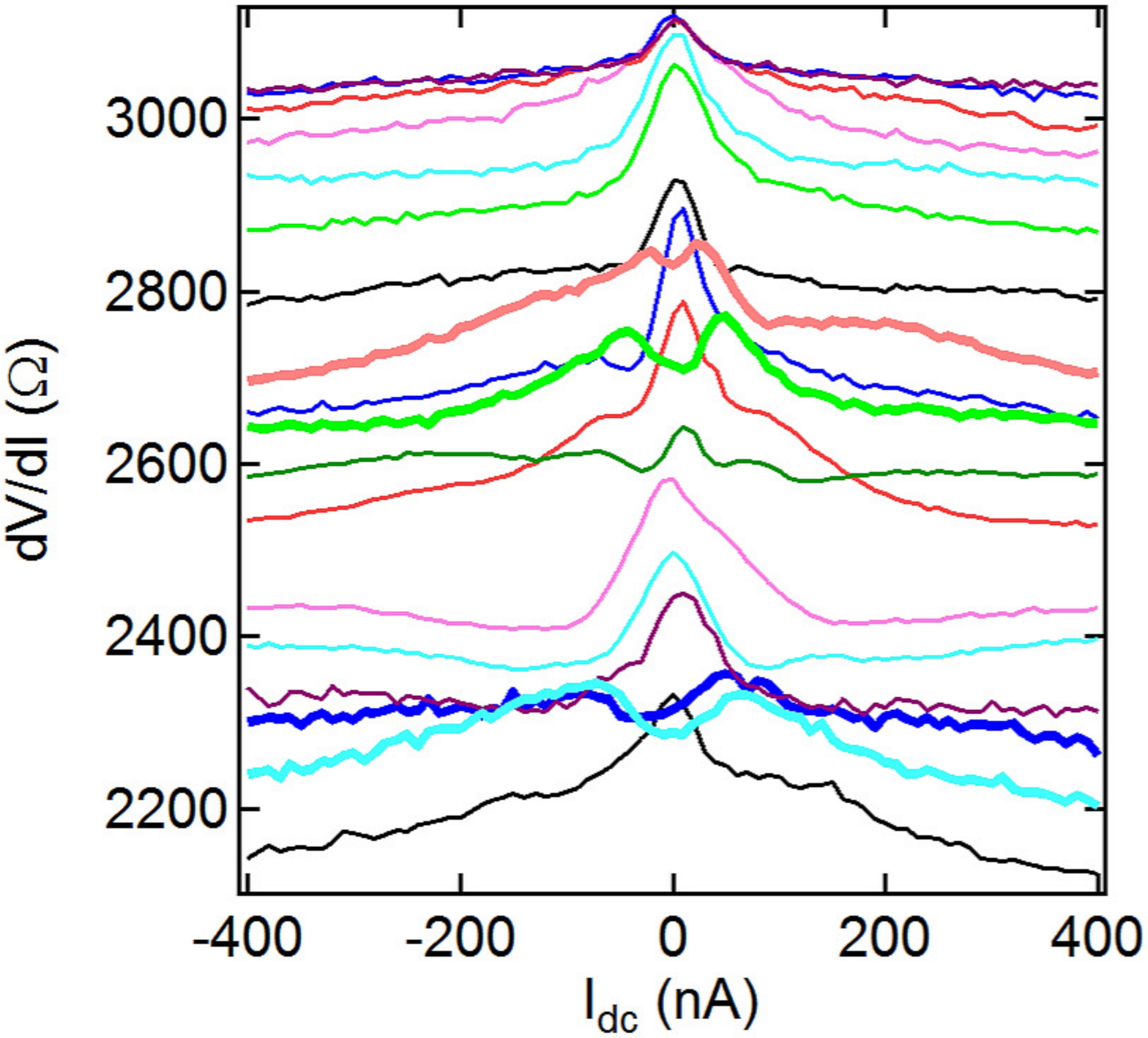}
\end{center}
\caption{Differential resistance versus dc current at selected gate voltages between -7 and 4 V at 7.5 T at 70 mK in the sample with ReW electrodes. The thicker curves are those which display a negative differential resistance at zero current, indicative of a superconducting proximity effect in the quantum Hall regime. Curves are not offset vertically.}
\label{dVdI_7p5T} 
\end{figure}

\begin{figure} 
\begin{center} 
\includegraphics[clip=true,width=10cm]{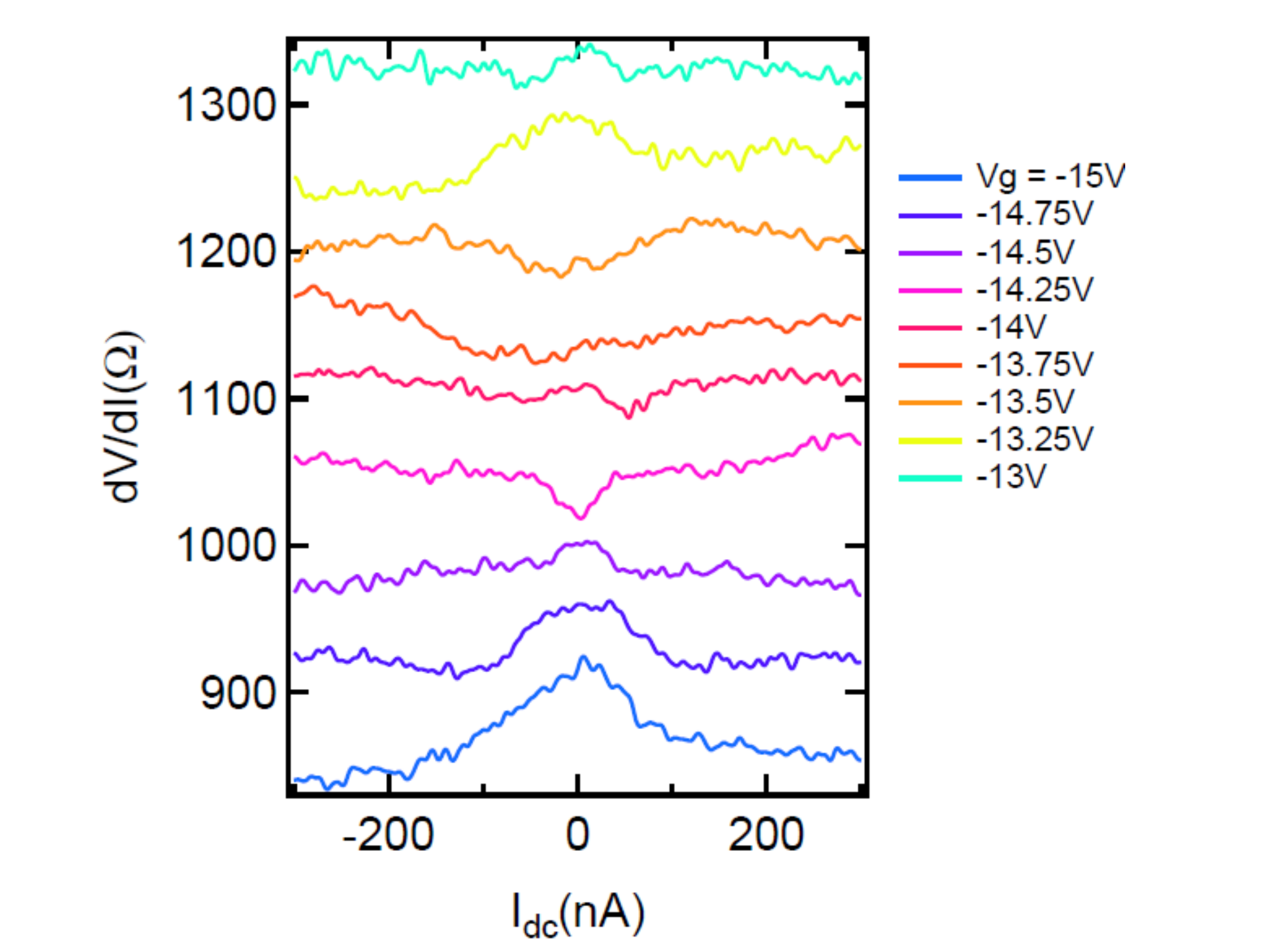}
\end{center}
\caption{Differential resistance versus dc current at 7.5 T, for gate voltages regularly distributed between -13~V and -15~V, in the sample with ReW electrodes. The alternation from dipped to peaked differential resistance at zero bias, with a 10\% variation, confirms that the oscillation in the $\nu=-6$ Hall plateau (seen in Fig. \ref{QHE_ReW}) is due to the proximity effect. Curves have been offset vertically for clarity.}
\label{dVdI_Vgsweep} 
\end{figure}

\begin{figure} 
\begin{center} 
\includegraphics[clip=true,width=8 cm]{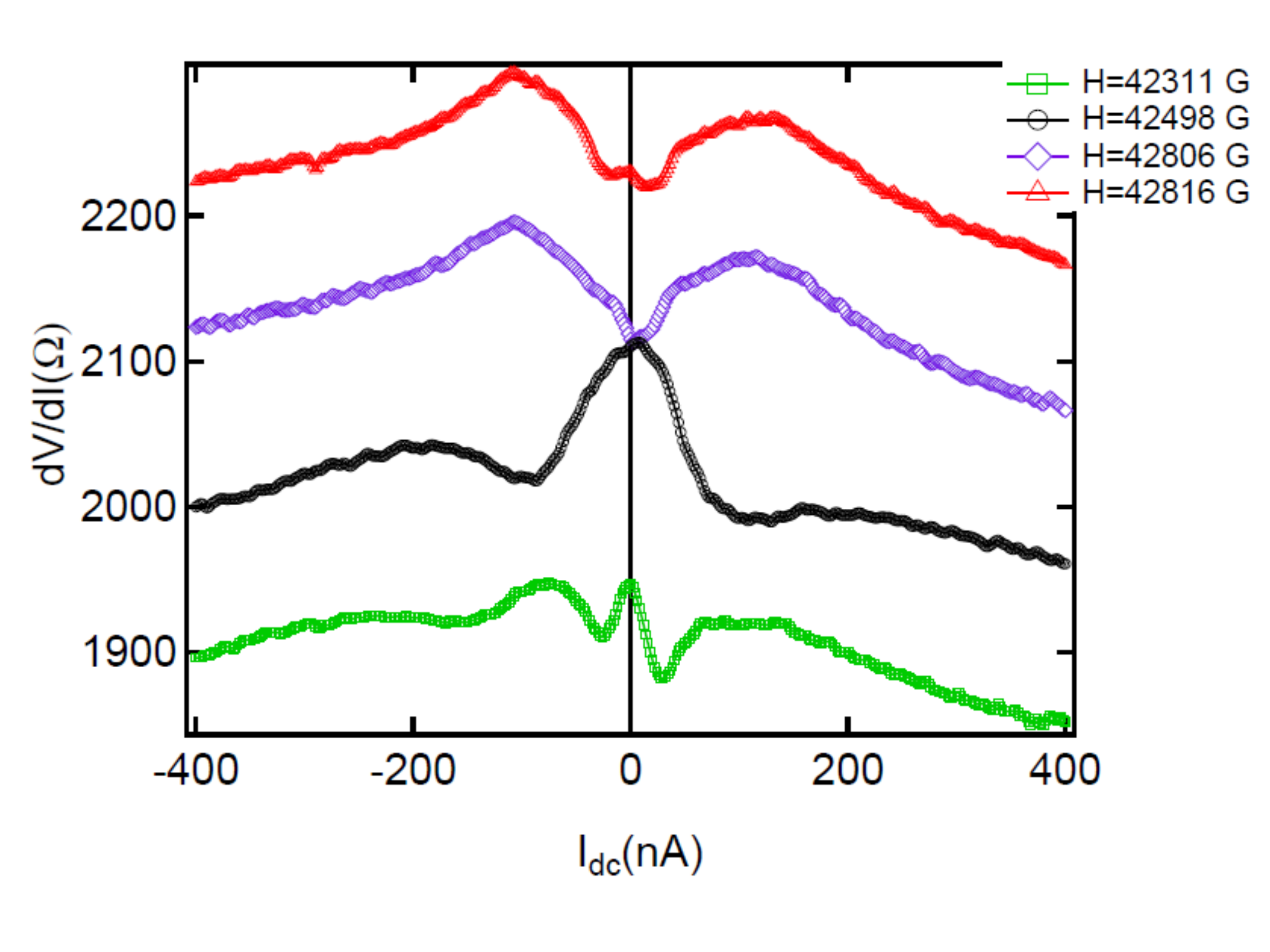}
\end{center}
\caption{Differential resistance as a function of dc current for selected curves at slightly different magnetic fields, around 4.2 T, at 55 mK and $V_g=0$, for the sample with ReW electrodes. The zero current differential resistance alternates between a peak and a dip, signaling the alternating nature of interference between transmitted Andreev pairs. The curves are offset vertically by $100~\Omega$ for clarity.}
\label{dip} 
\end{figure}

Signatures of a coherent proximity effect (i.e. a coherent propagation of pairs and a supercurrent) in the quantum Hall regime are visible when one exploits the non linearity of the reproducible fluctuations in the conductance (or resistance) as a function of magnetic field or gate voltage. These fluctuations, which stem from quantum interference between different diffusive trajectories, are known to be amplified in the case of superconducting contacts \cite{BeenakkerUCF}. But in some instances in this sample we find that the interference leads to a decrease of differential resistance around zero current, in contrast to the peaked differential resistance at zero current that is commonly observed in disordered samples at low temperature (due to electron-electron interactions or to the effect of the electromagnetic environment \cite{AA,DCB}).  Since in our two wire geometry the quantized Hall resistivity  adds to the zero longitudinal resistance of a supercurrent, we do not expect a zero two-wire resistance. But the signature of the supercurrent should be the differential resistance dip at zero bias. In addition, it was predicted in \cite{Ma} that the supercurrent intensity should be modulated by the Fermi energy or the magnetic field, in an Aharonov Bohm- like way. And, interestingly, we do observe alternating constructive and destructive interference, as a function of changing gate voltage or magnetic field, are demonstrated in Figs.  \ref{dVdI_7p5T}, \ref{dVdI_Vgsweep}, and \ref {dip}.  Similar features have been reported in 2D electron gases made in heterostructures in \cite{Takayanagi_dVdI} with varying magnetic fields,  but not gate voltages, and in samples in which no supercurrent was demonstrated at low field, in contrast to what we have achieved (see section II). In fact, we find that the dips in the differential resistance  have an amplitude of up to $50~\Omega$, and a current range of about $100~nA$ (see Figs. \ref{dVdI_7p5T}, \ref{dVdI_Vgsweep}, and \ref{dip}), comparable to the critical current measured in zero field. 

\begin{figure} 
\begin{center} 
\includegraphics[clip=true,width=10cm]{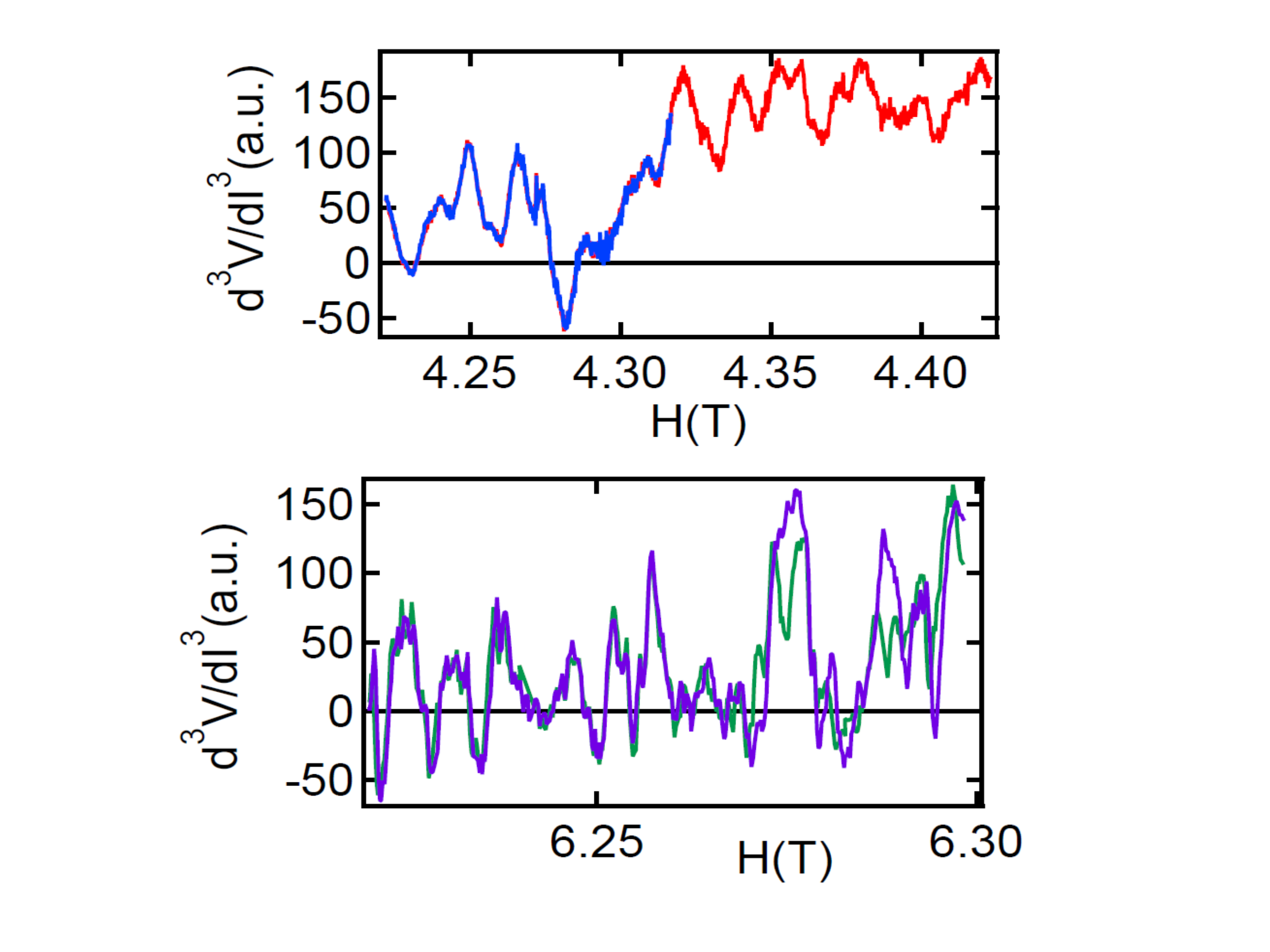}
\end{center}
\caption{Third derivative as a function of magnetic field at $V_g=0$ and T=55 mK, around 4.3 and 6.25 T. Increasing and decreasing field sweep directions are shown to demonstrate the reproducibility of the curves. A negative third derivative corresponds to a dipped differential resistance near zero bias, indicative of a superconducting proximity effect in the quantum Hall regime. Such a dipped differential resistance curve, alternating with peaked ones, is shown in Fig. \ref{dip}.}
\label{fluctuations_H} 
\end{figure}

Fig. \ref{fluctuations_H} illustrates how the low bias curvature of the differential resistance Vs current curves alternates in sign as the magnetic field is swept: the third derivative of the voltage Vs current curve is negative if the differential resistance is dipped at zero bias (induced proximity effect), but positive if the differential resistance is peaked (because of destructive interference, disorder, interactions). The oscillations, reminiscent of mesoscopic fluctuations, are reproducible and can be characterized by a correlation field $B_c\approx 100~G$ which varies with magnetic field and ac current excitation.


The fact that we find signatures of supercurrent (dips in the zero bias differential resistance) at low gate voltages (from -7 to +4V, see Fig. \ref{dVdI_7p5T}), for which the supercurrent was much reduced in zero field (see Fig. \ref{ReW_Scurrent}), points to the radically different effect of charge puddles at low and high magnetic field: we argued above that in zero field charge puddles tend to destroy the supercurrent because specular reflexion at the boundary between two oppositely charged puddles separates the two members of the Andreev pair, leading to large phase accumulation within each Andreev pair. The averaging of the many diffusing Andreev pairs leads to destructive interference. In contrast, in the quantum Hall regime, conduction proceeds in a ballistic-like fashion via a small number of channels (the edge states). If an edge state encounters a puddle boundary, it has been shown \cite{BeenakkerRMP,Marcus,Ullmo} that edge transport can proceed via an \lq\lq{}ambipolar snake state\rq\rq{}, made up of cyclotronic propagation with opposite rotations in the p and n regions.  Such edge states may contain a phase that depends on the specific disorder and puddle configuration at each edge, but the total accumulated dephasing over the round trip between the two superconducting electrodes should not average to zero, at low filling when only few edge channels propagate. Therefore the tuning of interference and thus of the proximity effect in the quantum Hall regime is expected over a larger gate voltage range than in the low field diffusive transport case.

\section{Conclusion}
In conclusion, we have shown that a proximity effect can be induced in a graphene junction up to 1.2 micrometers long. We find a stong suppression of the supercurrent near the charge neutrality point, and attribute it to the specular Andreev reflexion specific to monolayer graphene, at the boundaries between p and n puddles. This effect is all the stronger that the superconducting coherence length is short and that the junction is long, since Andreev pairs cannot avoid these junction regions.
In the Quantum Hall regime, a two wire measurement cannot reveal directly a supercurrent carried by edge states. But we argue that the dip in differential resistance at zero current is a signature of a supercurrent, flowing through the graphene via edge states which interfere constructively. This interference is modulated by gate voltage and magnetic field, as expected theoretically.
The question that needs to be addressed in the future is how to demonstrate that a supercurrent is circulating in the structure in the quantum Hall regime. Since a two wire transport measurement necessarily displays non zero resistance, one must find a different experimental configuration. In addition, it will be necessary to devise a method of distinguishing the dissipation-less supercurrent from the dissipation-less edge state transport. The detection of an orbital magnetic moment with a signature of pairs (via its field periodicity) \cite{magnet} may be a route towards this fascinating goal.

\section*{Acknowledgements}
We acknowledge H. Raffy for suggesting ReW as a superconducting material and the use of her cryostat, and B. Altshuler, J. Cayssol, B. Dassonneville, R. Deblock, V. Falko, M. Ferrier, J-N. Fuchs, M. O. Goerbig, A. Kasumov, and L. K. Lim for discussions.

\section*{Appendix}
\subsection{Quantum Hall effect in two samples with Nb electrodes and different aspect ratios}
\begin{figure} 
\begin{center} 
\includegraphics[clip=true,width=8cm]{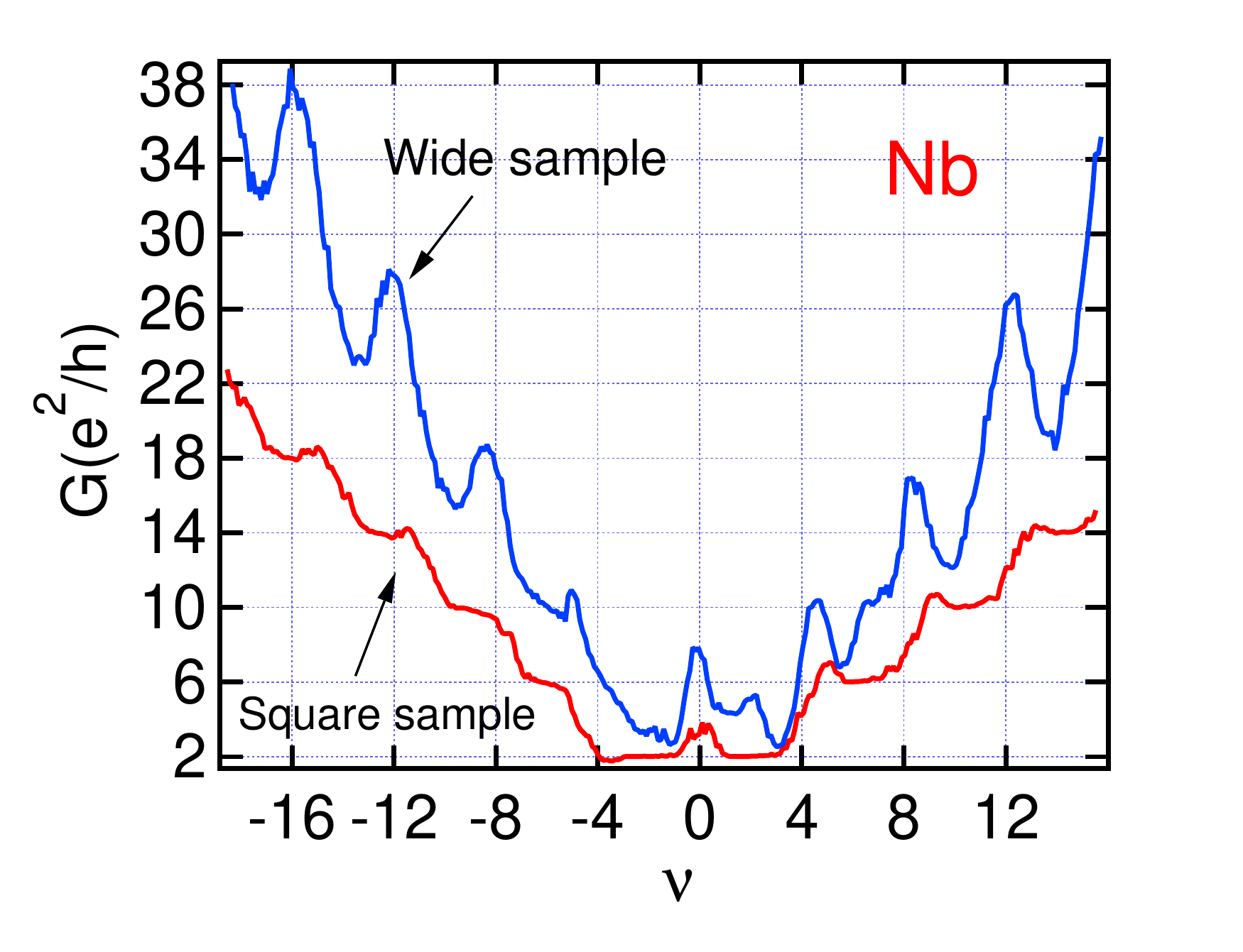}
\end{center}
\caption{Quantum Hall effect in two samples with Nb electrodes: the wide sample of Fig.\ref{photo} which displays a supercurrent at low field, and a square sample, that displayed no full proximity effect, but whose Hall quantization is closer to that expected for graphene. The data shows the two wire differential resistance as a function of filling factor, measured at 5T.}
\label{QHE_Nb_both} 
\end{figure}

We show in Fig. \ref{QHE_Nb_both} how the quantization expected for a two wire measurement of monolayer graphene in the Quantum Hall effect is better verified in a square sample than a wide geometry.

\subsection{Superconductivity of the ReW electrodes}
\begin{figure} 
\begin{center} 
\includegraphics[clip=true,width=8 cm]{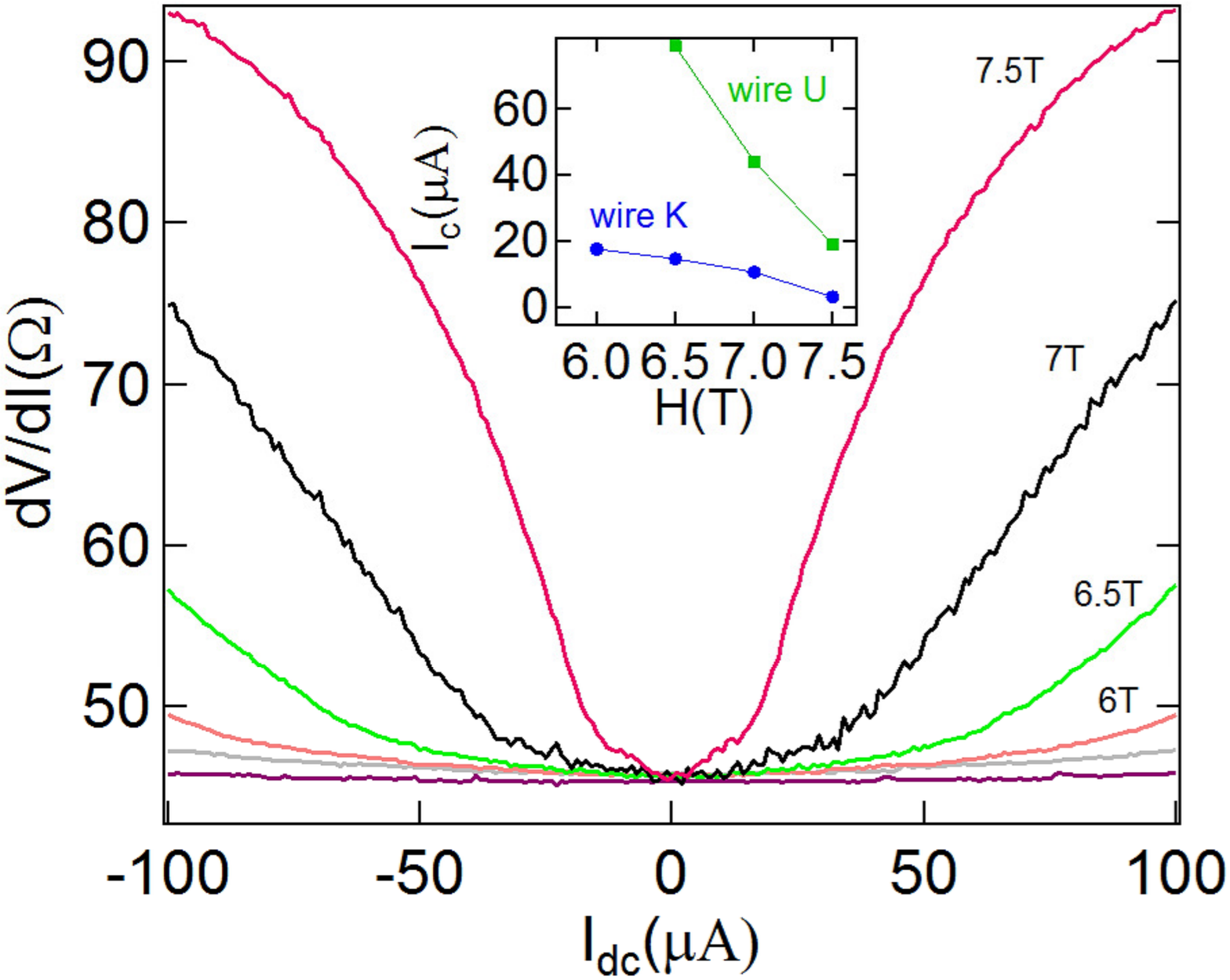}
\end{center}
\caption{Inset: Critical current as a function of magnetic field of two ReW leads that are slightly wider than the sample electrodes, at low temperature. Main panel, two wire differential resistance of one of the wire at different magnetic fields.}
\label{ReWwire} 
\end{figure}

Although we could not test the critical field of the electrode portion lying directly in contact with the graphene, we measured the critical current as a function of magnetic field of slightly wider ReW wires, and found that the critical current was larger than $3~\mu A$ at 55 mK, see Fig. \ref{ReWwire}.

\end{document}